\PassOptionsToPackage{table,xcdraw,x11names}{xcolor}
\documentclass[sigconf]{acmart}

\usepackage{booktabs}
\usepackage{graphicx}
\usepackage{xspace}
\usepackage{tikz}
\usepackage{pgfplots}
\usepackage{colortbl}
\usepackage{mdframed}
\usepackage{fontawesome}
\usepackage{fancybox}
\usepackage{array}
\usepackage{caption}
\usepackage{subcaption}

\pgfplotsset{compat=1.9}

\newcommand{\legendbox}[2]{%
    \begin{tikzpicture}[xscale=0.25,yscale=0.25]
        \draw [fill=#2!#1] (0,0) -- (1,0) -- (1,1) -- (0,1) -- cycle;
    \end{tikzpicture}
}

\newcommand{\UH}{\textsc{UH}\xspace}
\newcommand{\IIT}{\textsc{IITK}\xspace}
\newcommand{\etal}{\textit{et al.}\xspace}

\newcounter{observation}

\newcommand{\aigen}{GenAI\xspace}
\newcommand{\genai}{GenAI\xspace}
\newcommand{\genAI}{GenAI\xspace}

\newcolumntype{H}{>{\setbox0=\hbox\bgroup}c<{\egroup}@{}}
\newcommand{\SPACE}[1]{}

\newcommand{\todo}[1]{}
\renewcommand{\todo}[1]{{\color{red} TODO: {#1}}}

\title{Trust in Generative AI among Students}
\subtitle{An exploratory study}

\author{Matin Amoozadeh}
\affiliation{ 
  \institution{University of Houston}
 \city{Houston} 
  \state{TX} 
  \country{US}
}

\author{David Daniels}
\affiliation{ 
  \institution{Simon Fraser University}
  \city{Vancouver} 
  \state{BC} 
   \country{Canada}
}
\author{Daye Nam}
\affiliation{ 
  \institution{Carnegie Mellon University}
  \city{Pittsburgh} 
  \state{PA} 
   \country{US}
}

\author{Aayush Kumar}
\affiliation{ 
  \institution{Indian Institute of Technology}
  \city{Kanpur} 
  \state{} 
   \country{India}
}
\author{Stella Chen}
\affiliation{ 
  \institution{University of Houston}
  \city{Houston} 
  \state{TX} 
  \country{US}
}

\author{Michael Hilton}
\affiliation{ 
  \institution{Carnegie Mellon University}
  \city{Pittsburgh} 
  \state{PA} 
  \country{US}
}
\author{Sruti Srinivasa Ragavan}
\affiliation{ 
  \institution{Indian Institute of Technology}
  \city{Kanpur} 
  \country{India}
}
\author{Mohammad Amin Alipour}
\affiliation{
  \institution{University of Houston}
  \city{Houston} 
  \state{TX} 
   \country{US}  
}

\renewcommand{\todo}[1]{\textcolor{red}{#1}}
\begin{document}

\renewcommand{\shortauthors}{Matin Amoozadeh et al.}

\begin{abstract}

Generative Artificial Intelligence (\genai) systems have experienced exponential growth in the last couple of years. These systems offer exciting capabilities for CS Education (CSEd), such as generating programs, that students can well utilize for their learning.
Among the many dimensions that might affect the effective adoption of \genai for CSEd, in this paper, we investigate students' \textit{trust}. Trust in \aigen influences the extent to which students adopt \genai, in turn affecting their learning. In this paper, we present results from a survey of 253 students at two large universities to understand how much they trust \genai tools and their feedback on how \genai impacts their performance in CS courses.
Our results show that students have different levels of trust in \genai. We also observe different levels of confidence and motivation, highlighting the need for further understanding of factors impacting trust.

\end{abstract}



\maketitle

\section{Introduction}


Large Language Models (LLMs) and Generative Artificial Intelligence (\genai) have grown exponentially in the last two years and found interesting applications in various domains, such as in composing essays, developing question-answering chatbots, and code generation from incomplete text and code~\cite{wang2023investigating}.
They are also transforming the way users interact with computer systems: the use of natural text and dialog is more intuitive and appealing to more users of computers. In fact, \genai is expected to change the help-seeking behaviors and experiences of users ~\cite{kabir2023answers} and transform information search and discovery ~\cite{lund2023chatting}.

\genai is also beginning to influence computer science education (CSEd)~\cite{Berendt2020AIIE}.
Several studies have discussed and developed use cases for \genai in CSEd, especially as assistants that generate code examples or explanations for programming courses~\cite{DBLP:conf/sigcse/BeckerDFLPS23,DBLP:conf/ace/Finnie-AnsleyDB22,DBLP:conf/iticse/0001DMSBKTH23}.
However, the use of \genai in CSEd also raises concerns of plagiarism, potential biases, inaccurate results, and the potential development of unfavorable habits among students~\cite {prather2023s}. Moreover, even though students can -and do- seek \genai help to generate or explain code, they do not have sufficient expertise to evaluate the generated code. 

At this juncture, one factor, namely students' trust in \aigen, becomes key to whether and how students use \genai's code output. Trust determines the adoption of \aigen systems and their appropriate use~\cite{jacovi2021formalizing}: whereas overtrust could lead to students blindly accepting the AI's suggestions, undertrust could lead to ineffective use or abandonment of potentially useful aids. However, it is unclear whether students, who are novice programmers, can appropriately calibrate their trust in \genai tools, and if so at what level and with what prerequisite knowledge.  It is important to answer these questions about trust before we adopt \genai for CSEd.

Indeed, as we will see in Section~\ref{sec:rw:trust_in_ai}), trust in AI and automation has been studied, but mostly with professionals. Trust in AI remains largely understudied in an educational context. In this paper, we attempt to fill this gap. Specifically, we conducted an exploratory survey with 253 students at two large universities across US and India. The goal of the survey was to understand the extent to which students use and and trust \genai and how they perceive \genai as impacting their performance in CS courses. 
We also characterize the relation between students' trust in \aigen and other factors, such as motivation and confidence in programming, that affect student performance in CS classrooms. Our specific research questions are:

\begin{itemize}
\item[RQ1:]
How extensively do students use \aigen and what are their perceived benefits and drawbacks?

\item[RQ2:]
How much do students trust \aigen tools?

\item[RQ3:]
Overall, how do students perceive \genai in programming?

\end{itemize}
 
Our key findings include that students reported varying levels of trust in \aigen with $\sim$16\% participants expressing distrust, 36\% remaining neutral, and 47\% participants reporting trust in \aigen.
We also found a statistically significant correlation between trust and self-reported improvements in student motivation and confidence in programming due to \genai.
Particularly noteworthy is that the correlation between trust in \genai and motivation is more significant among first-generation students compared to continuing-generation students. 

These results have implications for \aigen developers in building effective tools, and for educators in fostering an appropriate level of trust among students, guiding them effectively in their learning.
\vspace{-2mm}
\section{Related Work}
\subsection{Studies on \genai Programming Assistants}
The recent widespread deployment of \genai programming assistants has motivated a lot of research on the use of these tools. Researchers have conducted empirical studies~\cite{leinonen.2023, Hellas.2023, Sarsa.2022} to evaluate the quality of code and explanations \genai programming assistants can generate.
Others have explored the usefulness of LLM-based programming tools~\cite{ziegler.2022, vaithilingam.2022, barke.2023, kazemitabaar.2023, xu.2022, mozannar.2022, ross.2023} via user studies.
Notably, Vaithilingam \etal~\cite{vaithilingam.2022} compared the user experience of GitHub Copilot with traditional auto-complete, and found that programmers faced more frequent difficulties in completing tasks with Copilot, although there was no significant impact on task completion time. 
Barke \etal~\cite{barke.2023} took a step further to understand \emph{how} programmers interact with code-generating models with a grounded theory analysis, and identified two primary modes of interaction, namely acceleration when \genai was used to speed up code authoring in small logical units, and exploration where 
\genai served as a planning assistant by suggesting structure or API calls. 

While these studies have contributed to our understanding of how developers use \genai developer tools, they primarily focused on evaluating the effectiveness of such tools and did not directly address developers' perceptions of them. 
Some studies (e.g., ~\cite{liang.2023, ziegler.2022}) have also focused on how developers perceive these tools.
For instance, Liang et al.~\cite{liang.2023} conducted a survey involving 410 developers to investigate the usability challenges associated with AI programming assistants, finding that although developers appreciate their autocomplete capabilities, they often face challenges in controlling the quality of code generation and concern about potentially infringing on intellectual property.
Although these studies provide valuable insights on how users \textit{perceive} the \genai tools, the insights are only applicable to professional programmers.
In this work, we study how \textit{students}, having  different levels of programming experience, skills, and self-efficacy, react to these tools.

\subsection{Studies on AI in Education}
With the advance of \genai programming assistants, research has also focused on the application of AI tools in computing education. 
Researchers and practitioners have delved into the opportunities and challenges of utilizing AI tools in education~\cite{Matin:2023:icer:trust, Denny.2023, Wang.20231i, DBLP:conf/sigcse/BeckerDFLPS23, lau2023ban}. These investigations have raised crucial questions regarding the benefits \genai offers individuals without formal education, the impact of easily accessible solutions on current pedagogical approaches, and the risks of plagiarism, potential biases, and the development of poor habits among students.

In addition to the theoretical discourse, many efforts have been made to explore the opportunities in 
assisting students in generating code~\cite{DBLP:conf/ace/Finnie-AnsleyDB22}, explanations~\cite{DBLP:conf/iticse/0001DMSBKTH23}, and identifying issues in problematic code that students seek help with~\cite{Hellas.2023}. These investigations have demonstrated that LLMs might be able to provide reasonably good help to students, but the actual benefit can vary depending on the complexity of the tasks. Building on these findings, several researchers have begun exploring ways to enhance students' efficient utilization of AI tools, given that the quality of AI-generated outputs heavily relies on the prompts provided by users. Denny \etal~\cite{Denny.2023} successfully improved GitHub Copilot's performance in introductory programming exercises from approximately 50\% to 80\% by experimenting with alternative prompts.
Our study contributes to this literature by providing additional insights into how students perceive \genai programming assistants and the precautions to take before adopting them for CS education.

\subsection{Trust in AI}
\label{sec:rw:trust_in_ai}
Trust is one of the main factors affecting the interaction between AI and its users. It is defined as ``the user’s judgment or expectations about how the AI system can help when the user is in a situation of uncertainty or vulnerability''~\cite{10.1145/3476068, cheng2022would}.
Both lack of trust~\cite{o2019question} and blind trust~\cite{DBLP:conf/sp/PearceA0DK22, DBLP:journals/corr/abs-2211-03622} affect how users interact with an AI tool and introduce risks in the human-AI interaction.
Therefore, researchers have studied how general users of AI tools build their trust in AI, to design tools that users can appropriately trust.
For instance, Jacovi \etal~\cite{DBLP:conf/fat/JacoviMMG21} took inspiration from the interpersonal trust model defined by sociologists and informed notions of the causes behind Human-AI trust, the connection between trust and Explainable AI (xAI), and the evaluation of trust.
Cheng \etal~\cite{cheng2022would}
studied online communities and found that developers collectively make sense of AI tools using the experiences shared by community members and leverage community signals to evaluate AI suggestions.
However, despite our understanding of the general AI tools, we still have very little understanding of the user trust in the AI tools and the implications of it, as they are still relatively new.


\section{Methodology}

\subsection{Data Gathering: Survey Study}

To better understand the dynamics of students' trust in AI for programming tasks, we conducted a survey with undergraduate and graduate students in large public universities in the US (\UH) and India (\IIT). Conducting the study across two sites allowed us to gather insights from a diverse group of participants, and provided better generalizability for our conclusions, especially following recent evidences from Human-Computer Interaction (HCI) and psychology about the need for diversity in participants along Western, Educated, Industrialized, Rich and Democratic (or WEIRD) dimensions \cite{linxen2021weird}.


\subsubsection{Survey Questions}
The survey included a range of questions including open and close-ended questions, covering demographics, participants' confidence in programming, trust in AI (as operationalized in Korber's Trust in Automation survey \cite{korber2019theoretical}), and their experiences and opinions about using AI tools.

The survey questionnaire was divided into five sections.
\begin{itemize}
    \item The first section consisted of demographic questions and included a question about participants' experience with \genai in programming. This question categorized participants into two groups, namely users and non-users of \genai.
    
    \item The second section was dedicated to participants who had used \genai; we elicited feedback on their experience with \genai, including their trust in the system, how \genai's use impacted their motivation and confidence in programming.
    \item The third section was for participants who had heard about \genai but had not yet used it. Here, we elicited participants' interest and willingness to try \genai tools and their perceived trust in such tools.
    \item The fourth section was for participants who had not heard about \genai and had no experience using it, where they were asked about their general awareness and impressions of the technology as well as their trust in AI. 
    
    \item Finally, the survey ended with an open-ended question asking participants' opinions on using AI in programming. 
\end{itemize}
By organizing the survey into these distinct sections, we aimed to gather comprehensive data from participants with varying levels of experience and exposure to \genai tools (e.g., ChatGPT, Copilot).

The survey mostly contained multiple-choice questions and therefore our analysis was predominantly quantitative. 
For the open-ended question, we used open-coding ~\cite{corbin2014qualitative} to identify and summarize the themes in students' responses.

We first conducted the survey at \UH. Based on the insights generated from preliminary analysis of this data, we included two additional questions in the survey conducted at \IIT. The first question related to English-language barriers, since English is not the native language of students in India. The second question related to students' trust in AI with respect to task complexity.

\subsubsection{Pilot}
The goal of the pilot was to: 1) identify ambiguities in the survey and 2) to estimate the time to complete the survey. To this end, one of the authors conducted a pilot survey at a third university, administering the survey to a group of teaching assistants. Based on the pilot participants' feedback, we refined our survey questions before distributing the final survey. The data from the pilot study are not included in our analysis.

\subsubsection{Participants}
We conducted the survey with undergraduate and graduate students at the University of Houston in the US and Indian Institute of Technology, Kanpur in India, we call them \UH and \IIT respectively in the rest of the paper. Table~\ref{tbl:DEMOALL} lists the demographics of the participants from the two study sites.

\begin{table}
\setlength{\arrayrulewidth}{0.1mm}
\setlength{\tabcolsep}{6pt}
\begin{small}
\caption{An overview of the demographics for all participants}
\label{tbl:DEMOALL} 
    
\begin{tabular}{ |p{3cm}|r|r||r| }  \hline
\rowcolor{Snow3} {\textbf{Demographics}} & \UH & \IIT & Total \\ \hline

\rowcolor{LightBlue1} \multicolumn{4}{|l|}{\textbf{Gender}} \\ \hline
    Male & 100 & 110 & 210 \\
    Female & 27 & 12 & 39 \\
    Other & 2 & 2 & 4 \\
    \hline

\rowcolor{LightBlue1} \multicolumn{4}{|l|}
{\textbf{College Generation}} \\ \hline
    First-generation & 59 & 32 & 91 \\ 
    Continuing-generation & 70 & 92 & 162 \\ \hline
    

\rowcolor{LightBlue1} \multicolumn{4}{|l|}{\textbf{Student levels}} \\ \hline
    Undergraduate & {122} & {93} & {215} \\
    Post-baccalaureate & 5  & 1  & 6  \\
    Graduate & 1  & 31  & 32  \\ \hline
    
\end{tabular}
\end{small}
\vspace{-5mm}
\end{table}

\subsubsection{Recruitment}
To recruit participants, we employed two distinct approaches, at the two study sites. At \UH, we targeted CS students, because we are primarily concerned with the use of \genai for programming. We therefore recruited students from two courses, namely Operating Systems (OS) and CS2 at \UH. 

At \IIT, our goal was to generalize the findings from \UH. We therefore targeted a diverse range of participants (from CS as well as non-CS backgrounds). We distributed the survey via the student emailing list containing over 6000 students, and received 110 responses (Table~\ref{tbl:DEMOALL}). The low response-rate (<2\%) is attributable to the fact that the study was conducted in the Summer of 2023, when most students were away.

Note that in both \UH and \IIT, the survey was conducted on a voluntary basis with participants opting in to the study.

\subsubsection{Protocol}
 To ensure ethical compliance, the Informed Consent form for this research study was created in adherence to the guidelines set by the Institutional Review Board (IRB) at both institutions. The Informed Consent form provided participants with essential information about the study, such as its voluntary nature, the estimated time required (here, no more than 15 minutes), and the assurance of confidentiality. It emphasized that no identifiable information would be collected or stored, and participants were given the option to provide their email addresses separately for future study purposes. The form highlighted that there were no foreseeable risks associated with participating in the research and encouraged participants to ask any questions they had before deciding to participate. 

\subsection{Data Analysis }

\subsubsection{Quantitative}
We used descriptive statistics, comparison of distributions, and correlations to identify different factors relating to students' trust in AI. 


\subsubsection{Qualitative}

In the qualitative section of our survey, we posed an optional open-ended question to the participants: \textit{In general, how do you feel about AI systems in programming?} We received a total of 73 responses,which we then analyzed using thematic open coding. Specifically, we coded each of the responses into 15 distinct themes induced from the data itself. The themes were \textit{fear}, the \textit{belief that AI will not replace human programmers}, perceptions of AI as \textit{helpful}, feelings of \textit{distrust}, \textit{trust} in AI systems, recognition of \textit{AI's assistance} in coding tasks, support for AI in \textit{basic coding}, acknowledgment of AI's role in \textit{learning}, recognition of AI's \textit{potential}, considerations of increased \textit{productivity}, concerns about \textit{bias}, views of AI as a \textit{hindrance}, awareness of \textit{ethical} threats, and opinions on the need for \textit{monitoring} AI systems. We also had a separate category for responses that were ambiguous in nature. 

For reliability, the responses were coded collaboratively by two authors, assigning one or more categories to each response. With the identification of each new code, we reviewed all prior comments for possible adjustments.

Such thematic coding of responses allowed us to draw meaningful insights about students' perspectives and shed light on the complex range of attitudes and emotions surrounding AI systems in the context of programming.


\section{Results}

Recall that our research questions are:

\begin{itemize}
    \item[RQ1] How extensively do students use \genai?
    \item[RQ2] How much do students trust \genai?
    \item[RQ3] Overall, how do students perceive \genai for programming?
\end{itemize}

\vspace{-2mm}








\subsection{RQ1: Exposure and Adoption of \genai}

\SPACE{
\begin{list}{}{\leftmargin=1em}
    \item \textbf{1.} How much exposure to programming AI do students have?
    \item \textbf{2.} How often do students use programming AI?
\end{list}
}
        
To understand students' use of \genai, we gathered whether students had heard of. Table~\ref{tbl:usage} summarizes the related findings. As the table shows, the majority of students have used \genai, and there was \emph{no} student that had not heard about \genai. This is unsurprising, considering the media coverage of tools such as ChatGPT.

\begin{table}
\begin{small}

\caption{Students' usage of \genai tools.}
\vspace{-3mm}

    \setlength{\arrayrulewidth}{0.1mm}
    \setlength{\tabcolsep}{6pt}
    \begin{tabular}{ |p{3cm}|r|r||r| } \hline \hline
    \rowcolor{Snow3} {\textbf{School}} & \UH & \IIT & Total \\ \hline
    \rowcolor{LightBlue1} \multicolumn{4}{|l|}{\textbf{Use of AI}} \\ \hline
        Used & 101 & 115 & 216 \\
        Heard but did not use & 28 & 9 & 37 \\\hline
    \rowcolor{LightBlue1} \multicolumn{4}{|l|}{\textbf{Tasks}} \\ \hline
        To seek help  & 74 & 79 & 153 \\
        To understand code  & 84 & 66 & 150 \\
        Learn new CS concepts  & 64 & 47 & 111\\
        Modify existing code & 49 & 52 & 101 \\
        Write New code  & 36 & 56 & 92 \\
        Other  & 26 & 45 & 71 \\ \hline
    \end{tabular}
    \label{tbl:usage}
\end{small}
\end{table}


Among the participants who had used \genai tools, we also elicited what all tasks they used them for. Table~\ref{tbl:usage}, Section Tasks lists these tasks. (Note that this question was optional and allowed multiple choices and so the total number responses for each task in Table~\ref{tbl:usage} does not equal the sample size.) 

As the table suggests, students used \genai for both programming and non-programming tasks (such as writing essays, as captured by the "Other" option). Surprisingly, within programming contexts, students mostly turned to \genai to seek programming help and to understand code, rather than writing new code itself. One possible reason for the less frequent use of \genai for writing new code is that academic honor codes prohibit the use of such tools for classroom purposes. 

Nonetheless, the use of \genai to seek help is noteworthy, and indicative of a larger trend of \genai replacing traditional internet searches. Whereas prior research on programming explains that seeking relevant -or \textit{valuable}- information by programmers is a time-consuming, or \textit{costly}, activity \cite{ko2007information}, \genai provides a low-cost way for programmers to gain high-value information. The notion of \textit{information foraging} \cite{pirolli}, therefore, needs a revisit especially in the programming context, since the costs of foraging for relevant information might now be replaced by tedium of prompting and foraging for evidences about the accuracy of \genai's responses, or its explanations.



The adoption of \genai tools might also be influenced by various socio-cultural aspects, as listed in Table~\ref{tbl:howitstarted}. A Wilcoxon signed-rank test,  comparing the responses of participants that had used \genai (or users) vs. those that had not (or non-users), suggest that the users "have been encouraged" more than non-users to use \genai. The users also generally believed that "professionals use AI" compared to their non-user counterparts ($p < 0.05$ in both cases).
 Meanwhile, \emph{non-users tend to worry more about AI replacing programmers ($p <0.5$).}


\begin{table*}
    \centering
    \caption{
        \label{tab:exposure}
        Distributions students' responses to adoption questions between  user and non-users.
    }
    \vspace{-3mm}
    {\small \begin{tabular}
        {
            >{\raggedright\arraybackslash}p{5.5cm}%
            >{\raggedright\arraybackslash}p{5.2cm}%
            >{\raggedright\arraybackslash}p{5.2cm}%
        }
        \toprule
        \multicolumn{3}{r}{
            \legendbox{100}{Firebrick1} Strong Disagree \,
            \legendbox{50}{Firebrick1} Disagree \,
            \legendbox{30}{gray} Neutral \,
            \legendbox{30}{Green1} Agree \,
            \legendbox{100}{Green1} Strong Agree
        } \\
        \midrule

        \multicolumn{1}{l}{ \bf Survey Question} &
        \multicolumn{1}{c}{\bf GenAI user (n=193)} &
        \multicolumn{1}{c}{\bf GenAI non-user (n=36)}\\
        
        \midrule
            { \bf  I have been encouraged to use AI tools. } &  \cellcolor{white} \begin{tikzpicture}[xscale=5.2,yscale=0.25] \draw [fill=Firebrick1!100] (0,0) -- (0,1) -- (0.04145077720207254,1) -- (0.04145077720207254,0) -- cycle; \draw [fill=Firebrick1!50] (0.04145077720207254,0) -- (0.04145077720207254,1) -- (0.2227979274611399,1) -- (0.2227979274611399,0) -- cycle; \draw [fill=gray!30] (0.2227979274611399,0) -- (0.2227979274611399,1) -- (0.5492227979274611,1) -- (0.5492227979274611,0) -- cycle; \draw [fill=Green1!30] (0.5492227979274611,0) -- (0.5492227979274611,1) -- (0.9015544041450776,1) -- (0.9015544041450776,0) -- cycle; \draw [fill=Green1!100] (0.9015544041450776,0) -- (0.9015544041450776,1) -- (0.9999999999999999,1) -- (0.9999999999999999,0) -- cycle;  \end{tikzpicture} & \begin{tikzpicture}[xscale=5.2,yscale=0.25] \draw [fill=Firebrick1!100] (0,0) -- (0,1) -- (0.1388888888888889,1) -- (0.1388888888888889,0) -- cycle; \draw [fill=Firebrick1!50] (0.1388888888888889,0) -- (0.1388888888888889,1) -- (0.3888888888888889,1) -- (0.3888888888888889,0) -- cycle; \draw [fill=gray!30] (0.3888888888888889,0) -- (0.3888888888888889,1) -- (0.7777777777777778,1) -- (0.7777777777777778,0) -- cycle; \draw [fill=Green1!30] (0.7777777777777778,0) -- (0.7777777777777778,1) -- (0.9444444444444444,1) -- (0.9444444444444444,0) -- cycle; \draw [fill=Green1!100] (0.9444444444444444,0) -- (0.9444444444444444,1) -- (1.0,1) -- (1.0,0) -- cycle;  \end{tikzpicture} \\
\rowcolor{gray!20} { \bf  Professionals use AI tools for programmin...} &  \cellcolor{white} \begin{tikzpicture}[xscale=5.2,yscale=0.25] \draw [fill=Firebrick1!100] (0,0) -- (0,1) -- (0.02072538860103627,1) -- (0.02072538860103627,0) -- cycle; \draw [fill=Firebrick1!50] (0.02072538860103627,0) -- (0.02072538860103627,1) -- (0.13471502590673576,1) -- (0.13471502590673576,0) -- cycle; \draw [fill=gray!30] (0.13471502590673576,0) -- (0.13471502590673576,1) -- (0.49740932642487046,1) -- (0.49740932642487046,0) -- cycle; \draw [fill=Green1!30] (0.49740932642487046,0) -- (0.49740932642487046,1) -- (0.8808290155440415,1) -- (0.8808290155440415,0) -- cycle; \draw [fill=Green1!100] (0.8808290155440415,0) -- (0.8808290155440415,1) -- (1.0,1) -- (1.0,0) -- cycle;  \end{tikzpicture} & \cellcolor{white} \begin{tikzpicture}[xscale=5.2,yscale=0.25] \draw [fill=Firebrick1!100] (0,0) -- (0,1) -- (0.2777777777777778,1) -- (0.2777777777777778,0) -- cycle; \draw [fill=Firebrick1!50] (0.2777777777777778,0) -- (0.2777777777777778,1) -- (0.75,1) -- (0.75,0) -- cycle; \draw [fill=gray!30] (0.75,0) -- (0.75,1) -- (0.9722222222222222,1) -- (0.9722222222222222,0) -- cycle; \draw [fill=Green1!30] (0.9722222222222222,0) -- (0.9722222222222222,1) -- (1.0,1) -- (1.0,0) -- cycle;  \end{tikzpicture} \\
{ \bf  Other people I know are using AI tools fo...} &  \cellcolor{white} \begin{tikzpicture}[xscale=5.2,yscale=0.25] \draw [fill=Firebrick1!100] (0,0) -- (0,1) -- (0.010362694300518135,1) -- (0.010362694300518135,0) -- cycle; \draw [fill=Firebrick1!50] (0.010362694300518135,0) -- (0.010362694300518135,1) -- (0.02590673575129534,1) -- (0.02590673575129534,0) -- cycle; \draw [fill=gray!30] (0.02590673575129534,0) -- (0.02590673575129534,1) -- (0.20207253886010362,1) -- (0.20207253886010362,0) -- cycle; \draw [fill=Green1!30] (0.20207253886010362,0) -- (0.20207253886010362,1) -- (0.7461139896373057,1) -- (0.7461139896373057,0) -- cycle; \draw [fill=Green1!100] (0.7461139896373057,0) -- (0.7461139896373057,1) -- (1.0,1) -- (1.0,0) -- cycle;  \end{tikzpicture} &  \begin{tikzpicture}[xscale=5.2,yscale=0.25] \draw [fill=Firebrick1!100] (0,0) -- (0,1) -- (0.08333333333333333,1) -- (0.08333333333333333,0) -- cycle; \draw [fill=Firebrick1!50] (0.08333333333333333,0) -- (0.08333333333333333,1) -- (0.19444444444444442,1) -- (0.19444444444444442,0) -- cycle; \draw [fill=gray!30] (0.19444444444444442,0) -- (0.19444444444444442,1) -- (0.27777777777777773,1) -- (0.27777777777777773,0) -- cycle; \draw [fill=Green1!30] (0.27777777777777773,0) -- (0.27777777777777773,1) -- (0.8055555555555556,1) -- (0.8055555555555556,0) -- cycle; \draw [fill=Green1!100] (0.8055555555555556,0) -- (0.8055555555555556,1) -- (1.0,1) -- (1.0,0) -- cycle;  \end{tikzpicture} \\
\rowcolor{gray!20} { \bf  I worry that AI is going to replace progr...} &  \cellcolor{white} \begin{tikzpicture}[xscale=5.2,yscale=0.25] \draw [fill=Firebrick1!100] (0,0) -- (0,1) -- (0.14507772020725387,1) -- (0.14507772020725387,0) -- cycle; \draw [fill=Firebrick1!50] (0.14507772020725387,0) -- (0.14507772020725387,1) -- (0.4455958549222798,1) -- (0.4455958549222798,0) -- cycle; \draw [fill=gray!30] (0.4455958549222798,0) -- (0.4455958549222798,1) -- (0.6373056994818653,1) -- (0.6373056994818653,0) -- cycle; \draw [fill=Green1!30] (0.6373056994818653,0) -- (0.6373056994818653,1) -- (0.8601036269430052,1) -- (0.8601036269430052,0) -- cycle; \draw [fill=Green1!100] (0.8601036269430052,0) -- (0.8601036269430052,1) -- (1.0,1) -- (1.0,0) -- cycle;  \end{tikzpicture} & \cellcolor{white} \begin{tikzpicture}[xscale=5.2,yscale=0.25] \draw [fill=Firebrick1!100] (0,0) -- (0,1) -- (0.05555555555555555,1) -- (0.05555555555555555,0) -- cycle; \draw [fill=Firebrick1!50] (0.05555555555555555,0) -- (0.05555555555555555,1) -- (0.2777777777777778,1) -- (0.2777777777777778,0) -- cycle; \draw [fill=gray!30] (0.2777777777777778,0) -- (0.2777777777777778,1) -- (0.4722222222222222,1) -- (0.4722222222222222,0) -- cycle; \draw [fill=Green1!30] (0.4722222222222222,0) -- (0.4722222222222222,1) -- (0.8333333333333333,1) -- (0.8333333333333333,0) -- cycle; \draw [fill=Green1!100] (0.8333333333333333,0) -- (0.8333333333333333,1) -- (0.9999999999999999,1) -- (0.9999999999999999,0) -- cycle;  \end{tikzpicture} \\

        \bottomrule
    \end{tabular}}
        \label{tbl:howitstarted}
\end{table*}




\vspace{-3mm}

\subsection{RQ2: Students' trust in \genAI}
\label{sec:rq:trust}


To gauge participants' trust in \genAI, we adapted the ``trust in AI'' instruments \cite{korber2019theoretical} and asked participants to use a five-point likert to indicate how much they agree with the following statements, each corresponding to one dimension of trust in~\cite{korber2019theoretical}: .  

\begin{itemize}
    \item[S1:] I trust the systems' output.
    \item[S2:] The output the system produces is as good as that which a highly competent person could produce.
    \item[S3:] I know what will happen the next time I use the system because I understand how it behaves.
    \item[S4:] I believe the output of the the system even when I don't know for certain that it is correct.
    \item[S5:] I have a personal preference for using such AI systems for my tasks.
    \item[S6:] Overall, I trust the AI system I use.
\end{itemize}


\begin{table*}[htbp]
    \centering
    \caption{
        Distributions of trust in \genai
        \label{tab:trust}
    }
    \vspace{-3mm}
    {\small \begin{tabular}%
        {
            >{\raggedright\arraybackslash}p{6.4cm}%
            >{\raggedleft\arraybackslash}p{9.8cm}%
        }
        \toprule
        \multicolumn{2}{r}{
            \legendbox{100}{Firebrick1} Strong Disagree \,
            \legendbox{50}{Firebrick2} Disagree \,
            \legendbox{30}{gray} Neutral \,
            \legendbox{20}{Green2} Agree \,
            \legendbox{100}{Green1} Strong Agree
        } \\
        \midrule
        {\bf Survey Question} &
        \multicolumn{1}{c}{\bf Distributions (n=242)}\\
        \midrule
            { \bf  I trust the system's output } & \cellcolor{white} \begin{tikzpicture}[xscale=9.8,yscale=0.25] \draw [fill=Firebrick1!100] (0,0) -- (0,1) -- (0.049586776859504134,1) -- (0.049586776859504134,0) -- cycle; \draw [fill=Firebrick1!50] (0.049586776859504134,0) -- (0.049586776859504134,1) -- (0.2727272727272727,1) -- (0.2727272727272727,0) -- cycle; \draw [fill=gray!30] (0.2727272727272727,0) -- (0.2727272727272727,1) -- (0.6322314049586777,1) -- (0.6322314049586777,0) -- cycle; \draw [fill=Green1!30] (0.6322314049586777,0) -- (0.6322314049586777,1) -- (0.975206611570248,1) -- (0.975206611570248,0) -- cycle; \draw [fill=Green1!100] (0.975206611570248,0) -- (0.975206611570248,1) -- (1.0,1) -- (1.0,0) -- cycle;  \end{tikzpicture} \\
\rowcolor{gray!20} { \bf  The output the system produces is as good...} & \cellcolor{white} \begin{tikzpicture}[xscale=9.8,yscale=0.25] \draw [fill=Firebrick1!100] (0,0) -- (0,1) -- (0.1446280991735537,1) -- (0.1446280991735537,0) -- cycle; \draw [fill=Firebrick1!50] (0.1446280991735537,0) -- (0.1446280991735537,1) -- (0.4132231404958677,1) -- (0.4132231404958677,0) -- cycle; \draw [fill=gray!30] (0.4132231404958677,0) -- (0.4132231404958677,1) -- (0.6900826446280992,1) -- (0.6900826446280992,0) -- cycle; \draw [fill=Green1!30] (0.6900826446280992,0) -- (0.6900826446280992,1) -- (0.9545454545454546,1) -- (0.9545454545454546,0) -- cycle; \draw [fill=Green1!100] (0.9545454545454546,0) -- (0.9545454545454546,1) -- (1.0,1) -- (1.0,0) -- cycle;  \end{tikzpicture} \\
{ \bf  I know what will happen the next time I u...} & \cellcolor{white} \begin{tikzpicture}[xscale=9.8,yscale=0.25] \draw [fill=Firebrick1!100] (0,0) -- (0,1) -- (0.04132231404958678,1) -- (0.04132231404958678,0) -- cycle; \draw [fill=Firebrick1!50] (0.04132231404958678,0) -- (0.04132231404958678,1) -- (0.15702479338842976,1) -- (0.15702479338842976,0) -- cycle; \draw [fill=gray!30] (0.15702479338842976,0) -- (0.15702479338842976,1) -- (0.4545454545454546,1) -- (0.4545454545454546,0) -- cycle; \draw [fill=Green1!30] (0.4545454545454546,0) -- (0.4545454545454546,1) -- (0.9049586776859504,1) -- (0.9049586776859504,0) -- cycle; \draw [fill=Green1!100] (0.9049586776859504,0) -- (0.9049586776859504,1) -- (1.0,1) -- (1.0,0) -- cycle;  \end{tikzpicture} \\
\rowcolor{gray!20} { \bf  I believe the output of the system ev...} & \cellcolor{white} \begin{tikzpicture}[xscale=9.8,yscale=0.25] \draw [fill=Firebrick1!100] (0,0) -- (0,1) -- (0.1487603305785124,1) -- (0.1487603305785124,0) -- cycle; \draw [fill=Firebrick1!50] (0.1487603305785124,0) -- (0.1487603305785124,1) -- (0.45867768595041325,1) -- (0.45867768595041325,0) -- cycle; \draw [fill=gray!30] (0.45867768595041325,0) -- (0.45867768595041325,1) -- (0.7727272727272727,1) -- (0.7727272727272727,0) -- cycle; \draw [fill=Green1!30] (0.7727272727272727,0) -- (0.7727272727272727,1) -- (0.9710743801652892,1) -- (0.9710743801652892,0) -- cycle; \draw [fill=Green1!100] (0.9710743801652892,0) -- (0.9710743801652892,1) -- (1.0,1) -- (1.0,0) -- cycle;  \end{tikzpicture} \\
{ \bf  I have a personal preference for using su...} & \cellcolor{white} \begin{tikzpicture}[xscale=9.8,yscale=0.25] \draw [fill=Firebrick1!100] (0,0) -- (0,1) -- (0.0743801652892562,1) -- (0.0743801652892562,0) -- cycle; \draw [fill=Firebrick1!50] (0.0743801652892562,0) -- (0.0743801652892562,1) -- (0.23553719008264462,1) -- (0.23553719008264462,0) -- cycle; \draw [fill=gray!30] (0.23553719008264462,0) -- (0.23553719008264462,1) -- (0.6198347107438016,1) -- (0.6198347107438016,0) -- cycle; \draw [fill=Green1!30] (0.6198347107438016,0) -- (0.6198347107438016,1) -- (0.9256198347107437,1) -- (0.9256198347107437,0) -- cycle; \draw [fill=Green1!100] (0.9256198347107437,0) -- (0.9256198347107437,1) -- (0.9999999999999999,1) -- (0.9999999999999999,0) -- cycle;  \end{tikzpicture} \\
\rowcolor{gray!20} { \bf  Overall, I trust the AI system I use. } & \cellcolor{white} \begin{tikzpicture}[xscale=9.8,yscale=0.25] \draw [fill=Firebrick1!100] (0,0) -- (0,1) -- (0.06198347107438017,1) -- (0.06198347107438017,0) -- cycle; \draw [fill=Firebrick1!50] (0.06198347107438017,0) -- (0.06198347107438017,1) -- (0.20661157024793386,1) -- (0.20661157024793386,0) -- cycle; \draw [fill=gray!30] (0.20661157024793386,0) -- (0.20661157024793386,1) -- (0.5661157024793388,1) -- (0.5661157024793388,0) -- cycle; \draw [fill=Green1!30] (0.5661157024793388,0) -- (0.5661157024793388,1) -- (0.9338842975206612,1) -- (0.9338842975206612,0) -- cycle; \draw [fill=Green1!100] (0.9338842975206612,0) -- (0.9338842975206612,1) -- (1.0,1) -- (1.0,0) -- cycle;  \end{tikzpicture} \\ 
        \bottomrule
    \end{tabular}}
    \vspace{-3mm}
    
\end{table*}

Table~\ref{tab:trust} shows the distribution of students' responses about trust along each of these dimensions.
Overall, students maintained a neutral view of \genai (average = 3.07, median=3). 
However, the trust does not seem to be blind: whereas 103 out of 216 participants disagreed with the statement ``I believe the output of the system even when I don't know for certain that it is correct'' (S5), only 45 agreed. 
These results generalized across gender, first-/continuing-generation students, and geographic location (p > 0.05 in Wilcoxon signed-rank tests). Here, first generation refers to students with both parents or guardians without a four-year college degree.

One point of difference across geographies is in where students place \genai vis-a-vis other expert programmers. Whereas students at \UH believed that "the output the system produces is as good as that which a highly competent person could produce", students at \IIT were more skeptical (mean likert scores were 3.02 vs. 2.64; $p < 0.05$ on Wilcoxon signed rank test). Surprisingly though \IIT students also expressed a slightly higher, but significant, personal preference in using AI than those that \UH (mean likert scores were 3.34 vs. 3.07 respectively; $p < 0.05$). Similar, but non-significant ($p=0.063$) differences were also observed between first-generation (mean=3.02) and continuing-generation (mean=2.71) students, with the former trusting \genai to produce output as good as a competent person. Future work must explore whether socio-economic factors (e.g., socio-economic status) underlie these observed differences in beliefs about \genAI's capabilities.



While analysis of aggregated data containing both \IIT and \UH, does not show any meaningful differences between gender, we found that, at \UH, the American  university, female users of \genai tend to trust \genai more than their male peers, on average,  3.31 vs. 3.07, for female and male students, respectively ($p < 0.05$). No such differences were found in the data from \IIT, calling for future work to replicate this result in other geographies and cultures and to explore whether and what cultural factors might contribute to such gender differences. 

\begin{figure*}
    \centering
    \begin{subfigure}[b]{0.3\textwidth}
        \centering
        \includegraphics[width=\textwidth,clip]{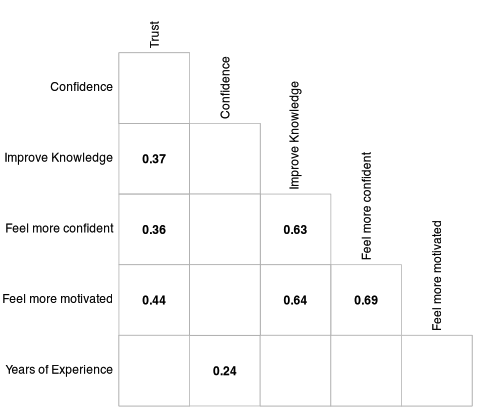}
    \vspace{-5mm}

         \caption{All Students}
    \label{fig:cor:all}
    \end{subfigure}
    \hfill
    \begin{subfigure}[b]{0.3\textwidth}
         \centering
        \includegraphics[width=\textwidth,clip]{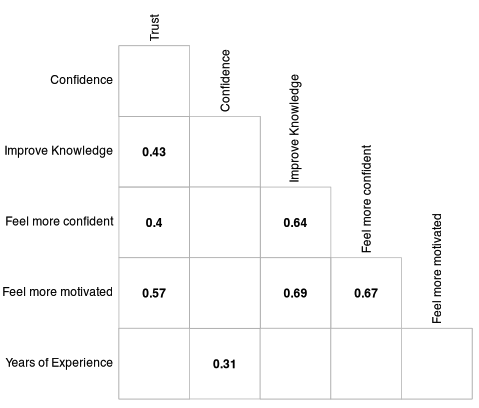}
        \vspace{-5mm}
         \caption{First-generation Students}
    \label{fig:cor:firstgen}
    \end{subfigure}
    \hfill
    \begin{subfigure}[b]{0.3\textwidth}
         \centering
        \includegraphics[width=\textwidth, clip]{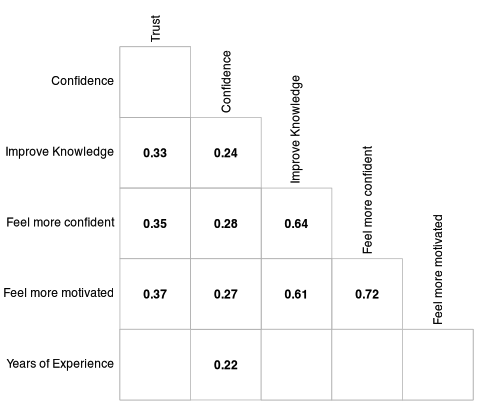}
        \vspace{-5mm}
         \caption{Continuing-generation Students}
    \label{fig:cor:cont}
     \end{subfigure}
             \vspace{-4mm}

     \caption{Correlation matrix between trust, confidence and motivational factors. The numbers in cells show correlations with $p-value \ll 0.01$. Empty cells represents no correlations. 
     }
    \vspace{-4mm}

    \label{fig:correlation}
\end{figure*}

At \UH, we also found that, participants from the lower level class, i.e., CS2 course, show more trust in \genai than those in the higher level class, i.e. OS: average likert scores are 3.23 vs. 2.95 for CS2 and OS students, respectively ($p < 0.05$). We note that students usually take OS at least a year after CS2. Unfortunately, we do not have year-wise comparable data from \IIT. However, as Figure~\ref{fig:correlation} shows, we do not see a correlation (Pearson, CI=95) between years of programming experience and perceptions of trust in \genai, in the overall data across both sites. Once again, the observation in \UH needs replication in other geographies and cultures.

Figure~\ref{fig:correlation} also lists correlations between students' trust in \genai and the degree to which \genai influences motivation and confidence. Overall (Figure~\ref{fig:cor:all}), the results suggest a moderate positive correlation between trust in \genai and the students' feelings of motivation and confidence when using those tools. 

Notably, among first-generation students (Figure~\ref{fig:cor:firstgen}),
trust correlates stronger to improving motivation ($0.57$) and confidence ($0.4$), compared to continuing-generation students, which are ($0.37$) and ($0.35$), respectively. In continuing-generation students, perception of confidence in programming was also moderately correlated with the degrees to which \genai help students to feel motivated and confident. The exact nature of causation is left for future work.

\subsection{RQ3: Student Perception of \genai}



Overall, 73 of participants from \UH and 41 from \IIT  answered the open-ended question about their general opinion about \aigen in programming. The top 3 themes pertinent to trust are as follows.

\subsubsection{Distrust (n=38)}
About half of participants expressed a lack of trust in the ability or output of \genai, and highlighted the need for human supervision and approval of AI-generated answers. They recognize that AI is not infallible and should be subject to human oversight to ensure accuracy and reliability. According to the respondents, this lack of trust stemmed from biases, errors, and limitations in AI algorithms. One respondent elaborated, \textit{"AI systems have a lot of potential in programming, as they can automate many repetitive and tedious tasks, make predictions based on large datasets, and even generate code. However, there are also concerns about the potential negative impacts of AI systems in programming. For example, there is the risk of bias and discrimination if the data used to train the AI system is not representative of the entire population."} Development of strategies to mitigate such concerns is essential to prevent any under-trust on \genai among students.


\subsubsection{Overall Helpfulness (n=24)} 
Around a quarter of respondents considered \genai as generally helpful. Although the specific tasks AI assists with were not specified, they acknowledged the positive impact of \genai. However, somewhat paradoxically, many students (especially those from advanced Computer Science courses) that found \genai helpful \textit{also} expressed distrust in the ability or output of AI. For instance, a student mentioned, \textit{"I use AI as a supplement, not a solution. I never 100\% trust the code it gives me, and often I have to modify it or tell the AI where or what it messed up on."}. Once again, careful considerations of these limitations of \genai and the role of trust in mediating or moderating the use of \genai is required before using these tools for CSEd.

\subsubsection{Contribution to Learning (n=14)} Finally, \genai learning tools can analyze and process vast amount of information, enabling them to identify patterns and provide relevant insights. Participants (N=14) believed that AI assists learning by enhancing concept understanding or perceiving the requirements of questions. One student noted: \textit{“I use ChatGPT when our classes give us new concepts to learn as well to explain code I give it. In these scenarios ... AI helped me learn the concepts a lot faster than if I were to try to watch videos ... I'd be able to take more advantage of it once I learn more about programming.”} These perceptions of \genai as more useful than traditional forms of learning (e.g., watching videos) have implications for both opportunities for \genai in both in-class and out-of-class learning for students.

\section{Threats to Validity}

\noindent{\textbf{Construct}}
\emph{Are we asking the right questions?}
We based our questions on previous work, notably operationalizing the notion of trust based on prior work. We also allowed students an "other" field so that they could provide answers we have not anticipated. We also iterated the questions based on insights from pilot studies. 

\noindent{\textbf{Internal}}
\emph{Did we skew the accuracy of our results with how we collected and analyzed data?}
Surveys by definition are affected by memory bias. We kept the surveys as short as possible to minimize fatigue, and randomized order of options wherever possible. We also used researcher triangulation in analyzing open-ended questions.

\noindent{\textbf{External}}
\emph{Do our results generalize?}
Because our survey has small sample size relative to overall student population, it is not possible to understand all student's perspective on \aigen.
To mitigate these concerns, we conducted the survey at two different universities, on two different continents, so as to get a global perspective. However, bias of drawing conclusions from predominantly STEM students and self-selection bias remain as threats to validity.


\section{Discussion and Conclusion}
\label{sec:discussion}
In this paper, we examined the perception of trust in \genai at two large public universities, each in the US and India.  
Our results suggest that \textit{all} students have heard about \genai tools and most have been using them, despite the fact they are relatively new. With the growing popularity and accessibility of these tools, we speculate that the adoption of \genai may just grow further, and more and more students will become regular users of \genai. 
Understanding how students trust such systems and what factors impact it can help us: 1) moderate their trust based on the actual capabilities of the systems, 2) design more trustworthy appropriate interfaces for education and 3) inform pedagogy that uses \genai in classrooms. 

Current \genai systems are based on large language models that are stubbornly opaque, despite active research in explainability and transparency of deep neural models~\cite{arrieta2020explainable}. It is difficult to reason about their behavior or comfortably predict their behavior in newly unseen settings. 
These systems exhibit emergent behaviors that their designers were not initially aware of~\cite{wei2022emergent}, and they can hallucinate and produce information-like output~\cite{bang2023multitask}.  
However, 118 out of 230 participants believed that \genai is transparent; that is, they believed that they "know what will happen the next time" and they can "understand how" \genai behaves (S3 in Section~\ref{sec:rq:trust}). 
The conflicting view between the lack of opacity of \genai models and the widespread students' belief that \genai is transparent, emphasizes the need for moderating trust in students for promoting the informed use of \genai among students.

GenAI tools are increasing in their capabilities and accuracy and it seems likely  that, in the near future, GenAI will be a tool that will be commonly used by most programmers. 
However, if students don't trust their tools, they will be unable to receive the benefits of using these tools. 
This could result in these students falling behind their peers who are able to use these GenAI tools.  Thus, CS educators should understand and identify the \emph{appropriate} level of trust that students should have in \genai tools.  Once this level of trust is identified, educators should find ways to help students calibrate their trust correctly.

Our results show variance in levels of trust in \genai in 
different demographics; for instance, we observed differences in trust between class standing and gender (at \UH).  It further highlights the need for further understanding of factors impacting trust in different populations. This is important for both educators to design exercises and pedagogy on how to introduce and encourage (or discourage) the use of \genai, and for designers of the \genai systems to consider in their models and interface design.

Our results suggest trust to be positively correlated with improving motivation, confidence, and knowledge in students regardless of their perception of confidence in their ability in programming. 
Interestingly, this correlation was stronger among first-generation students than continuing-generation students. 
It may present an opportunity for educators to encourage and retain more students in CS, or broadly STEM. Studies have shown that first-generation students feel insufficient access to support and lack a sense of belonging~\cite{terenzini1996first}, and graduate at a lower rate compared to continuing-generation students~\cite{deangelo2011completing}. Perhaps, \genai, if designed and used judiciously, can provide  extra support for the first-generation student and contribute to their progression and succession in CS programs.


\section*{Acknowledgments}
This material is based upon work supported by the U.S. National Science Foundation under Grant No. 2225373. Any opinions, findings, and conclusions or recommendations expressed in this material are those of the author(s) and do not necessarily reflect the views of the National Science Foundation.

\bibliographystyle{ACM-Reference-Format}
\bibliography{daye,bib,MatinPublications}

\end{document}